\documentclass[twocolumn,showpacs,amsfonts,aps,prc,superscriptaddress,nofootinbib,floatfix]{revtex4-1}
\usepackage{amsfonts}
\usepackage{amsthm}
\usepackage{bm}
\usepackage{graphicx}
\usepackage{amsmath}
\usepackage[usenames,dvipsnames]{color}
\usepackage{soul}
\usepackage{hyperref}
\usepackage{color}

\def\eq{{\,=\,}}

\begin{document}
\setstcolor{red}
\bibliographystyle{unsrt}

\title{Mode-coupling effects in anisotropic flow in heavy-ion collisions}

\author{Jing Qian}
\email[Correspond to\ ]{qianjing8758@gmail.com}
\affiliation{Department of Physics, Harbin Institute of Technology, Harbin, 150001, People's Republic of China} 
\affiliation{Department of Physics, The Ohio State University, Columbus, OH 43210-1117, USA}
\author{Ulrich Heinz}
\affiliation{Department of Physics, The Ohio State University, Columbus, OH 43210-1117, USA}
\author{Jia Liu}
\affiliation{Department of Physics, The Ohio State University, Columbus, OH 43210-1117, USA}
\begin{abstract}
Higher-order anisotropic flows in heavy-ion collisions are affected by nonlinear mode coupling effects. It has been suggested that the associated nonlinear hydrodynamic response coefficients probe the transport properties and are largely insensitive to the spectrum of initial density fluctuations of the medium created in these collisions. To test this suggestion, we explore nonlinear mode coupling effects in event-by-event viscous fluid dynamics, using two different models for the fluctuating initial density profiles, and compare the nonlinear coupling coefficients between the initial eccentricity vectors before hydrodynamic expansion and the final flow vectors after the expansion. For several mode coupling coefficients we find significant sensitivity to the initial fluctuation spectrum. They all exhibit strong sensitivity to the specific shear viscosity at freeze-out, but only weak dependence on the shear viscosity during hydrodynamic evolution.
\end{abstract}


\date{\today}

\maketitle

\section{Introduction}
\label{sec1}
\vspace*{-3mm}

Anisotropic flow, which is the hydrodynamic response to the anisotropic initial density profile, is one of the most important observables in relativistic heavy-ion collisions. It has been used to extract from experimental data on the transverse momentum distributions of hadrons emitted in the collision and their azimuthal angular correlations information about the transport coefficients of quark-gluon plasma (QGP), a novel state of strongly interacting matter that exists at extremely high temperature \cite{Heinz:2013th}.  The azi\-muthal asymmetry of the final state single-particle distribution is quantified by the complex anisotropic flow coefficients (``flow vectors'')
\begin{equation}
 V_n  \equiv v_ne^{in\Psi_n} \equiv \{e^{in\phi}\},
\label{Vdefi}
\end{equation}
where $\phi$ denotes the azimuthal angle around the beam direction of a particle emitted in the collision, $\{\dots\}$ denotes the average over all particles emitted in a given collision event, and $v_n$ and $\Psi_n$ denote the magnitude and azimuthal direction of the $n^\mathrm{th}$-order harmonic flow vector. The flow angle $\Psi_n$ can be written as $\Psi_n=\tilde\Psi_n+\Phi_\mathrm{RP}$ where $\Phi_\mathrm{RP}$ is the azimuthal angle of the reaction plane, spanned by the impact parameter and beam direction, and $\tilde\Psi_n$ denotes the direction of the $n^\mathrm{th}$-order flow relative to that plane and is thus directly affected by the impact parameter dependent collision geometry, especially in peripheral collisions. Since in the lab frame the reaction plane angle $\Phi_\mathrm{RP}$ is uniformly distributed in the event sample, the ensemble average 
\begin{equation}
\label{Vn}
   \langle V_n\rangle = 0
\end{equation}
for all $n$. Only combinations of $V_n$ that are independent of $\Phi_\mathrm{RP}$ can have non-vanishing ensemble expectation values, and only such combinations are considered in this work. 

In a similar spirit to Eq.~(\ref{Vdefi}), the azimuthal spatial anisotropies of the initial energy density profile $e(r,\varphi)$ in the transverse plane (which fluctuate from event to event due to quantum fluctuations of the positions of the nucleons inside the colliding nuclei and of the gluon fields that mediate the interactions between the colliding nuclei inside those nucleons at the time of impact) are usually characterized by complex eccentricity coefficients defined as the following energy density moments in the transverse plane (see e.g. \cite{Alver:2010dn,Teaney:2010vd,Petersen:2010cw,Qiu:2011iv}):
\begin{equation}
\mathcal{E}_n  \equiv \epsilon_n e^{in\Phi_n} \equiv 
  - \frac{\int d^2r_\perp\, r^m\,e^{in\varphi}\,e(r,\varphi)}
           {\int d^2r_\perp\, r^m\,e(r,\varphi)} .                                                      
\label{Edefi}
\end{equation}
We use $m\eq{n}$ for $n{\geq\,}2$ and $m\eq3$ for $n\eq1$ \cite{Teaney:2010vd}. The angle $\Phi_n$ in Eq.~(\ref{Edefi}) can be written as $\Phi_n\eq\tilde\Phi_n{+}\Phi_\mathrm{RP}$ where $\tilde\Phi_n$ is known as the $n^\mathrm{th}$-order participant plane angle relative to the reaction plane.

In theoretical simulations, initial conditions are usually created in the reaction plane frame. The corresponding theoretical eccentricity and flow coefficients $\tilde{\mathcal{E}}_n$ and $\tilde V_n$ thus have phase factors given by $e^{in\tilde\Phi_n}$ and $e^{in\tilde\Psi_n}$, respectively. To simulate the experimental situation, we can multiply all the theoretically computed $\tilde{\mathcal{E}}_n$ and $\tilde V_n$ coefficients for a given event by a random phase $e^{in\Phi_\mathrm{RP}}$ representing the random orientation of the reaction plane for this event, thereby ensuring that $\langle \mathcal{E}_n\rangle\eq\langle \tilde{\mathcal{E}}_n e^{in\Phi_\mathrm{RP}}\rangle\eq\langle V_n\rangle\eq\langle \tilde V_n e^{in\Phi_\mathrm{RP}}\rangle\eq0$, consistent with Eq.~(\ref{Vn}).  However, if (as done here) only combinations of eccentricity or flow combinations are studied in which the dependence on $\Phi_\mathrm{RP}$ cancels, this additional step is unnecessary, and we can directly substitute the theoretically computed coefficients in the reaction plane frame for the experimentally measured ones in the lab frame.

Theoretical calculations have shown that for elliptic and triangular flows ($n\eq2$, 3), the magnitudes of the anisotropic flow coefficients $v_n$ are approximately linear in the magnitude of the initial eccentricity coefficients $\epsilon_n$, except for large impact parameters \cite{Qiu:2011iv,Niemi:2012aj,Fu:2015wba,Niemi:2015qia,Noronha-Hostler:2015dbi}. Since shear viscosity reduces the hydrodynamic response $v_n$ to the initial eccentricity $\epsilon_n$, $v_n/\epsilon_n$ for $n\eq2$,\,3 was proposed in \cite{Song:2010mg,Qiu:2011iv} as a clean observable to quantitatively constrain the shear viscosity of quark-gluon plasma. Unfortunately, the initial eccentricities $\epsilon_n$ are not directly measurable and are plagued by significant model uncertainties, which lead to even larger uncertainties in the shear viscosities extracted from elliptic and triangular flow data \cite{Song:2010mg,Qiu:2011hf}. Higher-order $v_n$ harmonics are more sensitive to shear viscosity than elliptic and triangular flow \cite{Alver:2010dn,Schenke:2011bn}, which initially gave some hope that it might be possible to constrain both the shear viscosity and the initial eccentricity spectrum simultaneously by analyzing the full set of $v_n$ flow harmonics. However, the response of the higher-order $v_n$ coefficients to the corresponding initial eccentricities $\epsilon_n$ is nonlinear due to mode coupling \cite{Qiu:2011iv,Gardim:2011xv,Teaney:2012ke}, rendering the realization of this idea much less straightforward than first thought.  

Additional independent information on the initial eccentricity spectrum $\mathcal{E}_n$ and the transport properties of the expanding medium that converts these initial eccentricities into anisotropic flows in the final state is contained in correlations among the flow angles $\Psi_n$ (a.k.a. event-plane correlations) \cite{Jia:2012ma,Jia:2012sa,Qiu:2012uy,Bhalerao:2013ina,Teaney:2013dta}. Correlations among the anisotropic flow magnitudes $v_n$ were measured experimentally and shown to exhibit unmistakable evidence for nonlinear mode coupling during the dynamical evolution of the fireball \cite{Aad:2015lwa}. However, the main question whether it is possible to find observables that separate the sensitivity of the final complex flow coefficients $V_n$ to the transport properties of the evolving medium from that to the (experimentally not directly measurable) fluctuating initial density profiles remained unanswered by all of the above analyses.

In two interesting recent papers \cite{Bhalerao:2014xra,Yan:2015jma} Ollitrault and collaborators introduced a set of nonlinear hydrodynamic mode coupling coefficients (defined in Sec~\ref{sec2}) which, they suggested, should be independent of the fluctuating initial density profiles and hence a clean probe of the transport properties of the liquid medium. If true, this would open the door to measuring quark-gluon transport coefficients without being affected by model uncertainties for the initial eccentricity coefficients and their fluctuation spectra. Prescriptions for separating the nonlinear response from the linear terms were given in Refs.~\cite{Bhalerao:2014xra} and \cite{Jia:2014jca}. In this work we test both, these prescriptions and the initial-state independence of the nonlinear coupling coefficients, as well as the latter's sensitivity to the QGP shear viscosity, using event-by-event hydrodynamic simulations with fluctuating initial conditions from the Monte Carlo Glauber (MC-Glb) and Monte Carlo Kharzeev-Levin-Nardi (MC-KLN) models.  

This paper is organized as follows: In Section~\ref{sec2} we start by studying ``linear'' and ``nonlinear'' contributions to the higher-order harmonic flows $V_n$, following Yan and Ollitrault's prescription \cite{Yan:2015jma} of keeping only the largest nonlinear mode coupling terms involving at least one factor of $V_2$ and $V_3$ (mode coupling terms involving higher-order $V_n$ coefficients are expected to be smaller because of stronger shear viscous damping). A detailed discussion of the correlations between the so defined ``linear'' and ``nonlinear'' response contributions shows, however, that the ``linear'' contribution $V_{nL}$ defined by this decomposition cannot be identified with the linear response to the corresponding initial eccentricity $\mathcal{E}_n$,  $V_{nL}{\,\ne\,}\alpha_n\mathcal{E}_n$, contrary to what was previously thought \cite{Bhalerao:2014xra,Yan:2015jma}. After suitably reinterpreting the decomposition of the flows $V_n$, we use hydrodynamic simulations to calculate the mode coupling coefficients as well as the statistical correlations between the different terms in the decomposition. In Section~\ref{sec3} we generalize the analysis by introducing additional allowed mode coupling terms and study the behavior of their coefficients. Trying to trace the origin of the mode coupling in the anisotropic flows, we investigate in Section~\ref{sec4} correlators between the initial eccentricity coefficients that are defined in analogy with the nonlinear mode coupling coefficients for the final flows. Our conclusions are summarized in Section~\ref{sec6}. The absence of correlations between the leading and non-linear mode coupling terms discussed in Section~\ref{sec3} is checked in Appendix~\ref{appa}. Appendix~\ref{appb} contains a discussion of resonance decay effects on the mode coupling coefficients. 

All our hydrodynamic simulations are done for Pb+Pb collisions at $\sqrt{s}\eq2.76\,A$\,TeV, using the  {\sc iEBE-VISHNU} code package \cite{Shen:2014vra}. They start at $\tau_0\eq0.6$\,fm/$c$ without pre-equilibrium flow, end on an isothermal freeze-out surface with temperature $T_{\mathrm{dec}}\eq120$\,MeV, and use as default choices for the specific shear viscosity the values $\eta/s\eq0.08$ for MC-Glb initial conditions and $\eta/s\eq0.2$ for MC-KLN initial profiles. The initial conditions were obtained from an older version of {\sc iEBE-VISHNU} that does not account for multiplicity fluctuations in $pp$ collisions. The anisotropic flow coefficients are calculated on the freeze-out surface using the Cooper-Frye algorithm, including all important resonance decay contributions \cite{Qiu:2012tm} unless stated otherwise. For each initial condition model and each centrality bin, we performed 2000 hydrodynamic runs with fluctuating initial profiles.

\vspace*{-2mm}
\section{Mode coupling to $V_2$ and $V_3$}
\label{sec2}
\vspace*{-3mm}

%
\begin{figure}[b]
    \includegraphics[width=\linewidth]{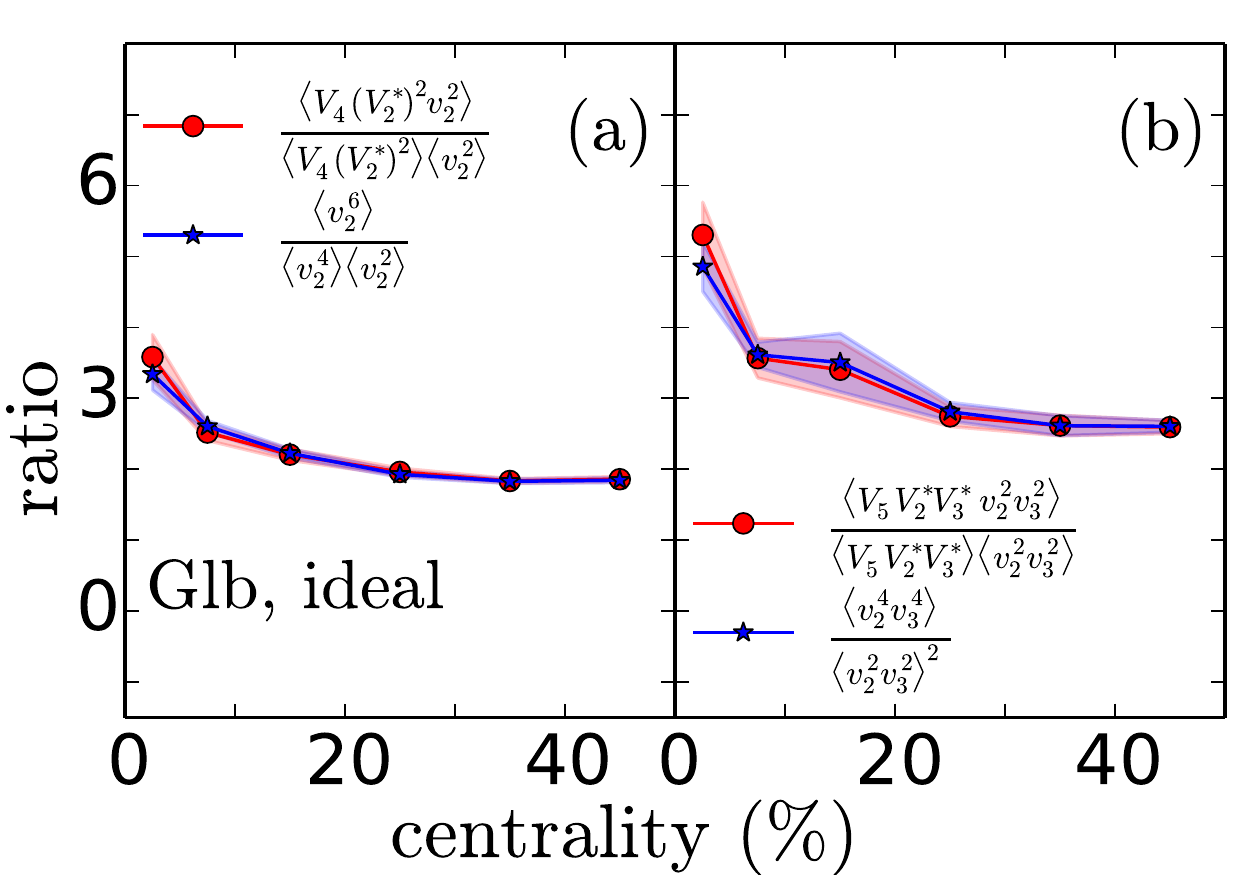}
    \caption{(Color online)
    Test of Eqs. (\ref{cor_VlVnl}). Full circles correspond to the left-hand sides, using ideal hydrodynamics 
    with MC-Glauber initial conditions for Pb-Pb collisions at 2.76\,$A$\,TeV.  Stars correspond to the
    right-hand sides. 
    \label{F1}
}
\end{figure}
%

In \cite{Teaney:2012ke,Yan:2015jma}, $V_4$ and higher harmonics are modeled as the sum of linear and nonlinear response terms, keeping only nonlinear terms involving the two largest anisotropic flow coefficients $V_2$ and $V_3$:\footnote{Higher-order anisotropies are suppressed by viscous suppression, and the directed flow $V_1$ is special because it is strongly constrained by transverse momentum conservation \cite{Luzum:2010fb}.}
\begin{equation}
\begin{split}
  V_4 & = V_{4 L} + \chi_{422} V_2^2,\\
  V_5 & = V_{5 L} + \chi_{523} V_2 V_3,\\
  V_6 & = V_{6 L} + \chi_{633} V_3^2 + \chi_{6222} V_2^3,\\
  V_7 & = V_{7 L} + \chi_{7223} V_2^2 V_3.
\end{split}
\label{VlVnl_old}
\end{equation}
In this decomposition, it was implicitly assumed that the linear parts $V_{nL}$ describe the linear response to the initial eccentricities of the same harmonic order while the nonlinear parts are a response to lower eccentricity harmonics. The authors of \cite{Yan:2015jma} therefore expected the correlation between the linear and nonlinear terms on the right-hand side of Eqs.~(\ref{VlVnl_old}) to be weak. This expectation makes the implicit assumption, however, that the initial eccentricities $\mathcal{E}_n$ are uncorrelated. Due to the almond-like geometric deformation of the transverse nuclear overlap zone in non-central collisions we expect instead that the participant plane angles of the average even-order eccentricities $\mathcal{E}_2$, $\mathcal{E}_4$, $\mathcal{E}_6$ are correlated with each other. Moreover, ATLAS \cite{Aad:2015lwa} observed significant anti-correlation between the magnitudes $v_2$ and $v_3$ in non-central Pb+Pb collisions at the LHC and showed that, due to geometric bias, similar anti-correlations exist already between the corresponding eccentricities $\varepsilon_2$ and $\varepsilon_3$ in the initial state unless the collisions are very central. These considerations prompted us to check the proposed \cite{Yan:2015jma} lack of correlation between the various terms on the right hand side of Eq.~(\ref{VlVnl_old}) using hydrodynamic simulations.

To perform this check we use a method proposed in Ref.~\cite{Bhalerao:2014xra}. Taking $V_4$ and $V_5$ as examples, if the linear and nonlinear parts are assumed to be uncorrelated, the following relations between moments of the $V_n$ distributions hold \cite{Bhalerao:2014xra}:
\begin{equation}
\begin{split}
\frac{\left\langle V_4 (V_2^*)^2 v_2^2 \right\rangle}
       {\left\langle V_4 (V_2^*)^2 \right\rangle \left\langle v_2^2 \right\rangle} & = 
\frac{\left \langle v_2^6 \right\rangle}
       {\left\langle v_2^4 \right\rangle \left\langle v_2^2 \right\rangle},\\
\frac{\langle V_5 V_2^* V_3^* v_2^2 v_3^2 \rangle}{\langle V_5 V_2^* V_3^* \rangle\, \langle v_2^2 v_3^2 \rangle} & = \frac{\langle v_2^4 v_3^4 \rangle}{\langle v_2^2 v_3^2 \rangle ^2}.
\end{split}
\label{cor_VlVnl}
\end{equation}
In Ref.~\cite{Bhalerao:2014xra} this assumption was tested and found to hold in the AMPT model, and Fig.~\ref{F1} shows that it also holds when the initial conditions are evolved hydrodynamically (in this case using ideal (inviscid) fluid dynamics). While in central collisions smaller $v_2$ values and relatively larger fluctuations cause larger statistical uncertainties for the ratios, the agreement between the left and right hand sides of Eqs.~(\ref{cor_VlVnl}) is found to be good at all collision centralities. Similar results were found for MC-KLN initial conditions and non-zero values of the shear viscosity (not shown).

Instead of assuming that the linear and nonlinear terms in Eqs.~(\ref{VlVnl_old}) are uncorrelated and testing this assumption via the relations (\ref{cor_VlVnl}), we could  try to directly compute the Pearson correlation coefficients between them \cite{Bhalerao:2014xra}. The Pearson correlation coefficient between two complex variables $f$ and $g$ with vanishing means, $\langle f \rangle\eq\langle g \rangle\eq0$, is defined as 
\begin{equation}
\mathrm{Cor}(f,g) = \frac{\langle f g^* \rangle}{\sqrt{\langle |f|^2 \rangle \langle |g|^2 \rangle }}.
\label{complex_correlation_simple}
\end{equation}
Fig.~\ref{F1} suggests that, if we could perform the separation (\ref{VlVnl_old}) in our calculations and calculated the Pearson correlation coefficients between the linear and nonlinear terms, we should find $\mathrm{Cor}(V_{4L},V_2^2){\,\approx\,}\mathrm{Cor}(V_{5L},V_2V_3){\,\approx\,}0$. Unfortunately, it is not known how to perform the separation (\ref{VlVnl_old}) event by event. However, we know from years of hydrodynamic simulations \cite{Qiu:2011iv,Niemi:2012aj,Fu:2015wba,Niemi:2015qia,Noronha-Hostler:2015dbi} that $V_2$ and $V_3$ are dominated by linear response to the initial ellipticity $\mathcal{E}_2$ and triangularity $\mathcal{E}_3$, respectively. If, as assumed in \cite{Bhalerao:2014xra,Yan:2015jma}, $V_{4L}$ and $V_{5L}$ describe similarly the linear response to $\mathcal{E}_4$ and $\mathcal{E}_5$, respectively, we should therefore expect their Pearson correlation coefficients to satisfy the identities $\mathrm{Cor}(V_{4L},V_2^2)\eq\mathrm{Cor}(\mathcal{E}_4,\mathcal{E}_2^2)$ and $\mathrm{Cor}(V_{5L},V_2V_3)\eq\mathrm{Cor}(\mathcal{E}_5,\mathcal{E}_2\mathcal{E}_3)$.

\begin{figure}[!htb]
    \includegraphics[width=\linewidth]{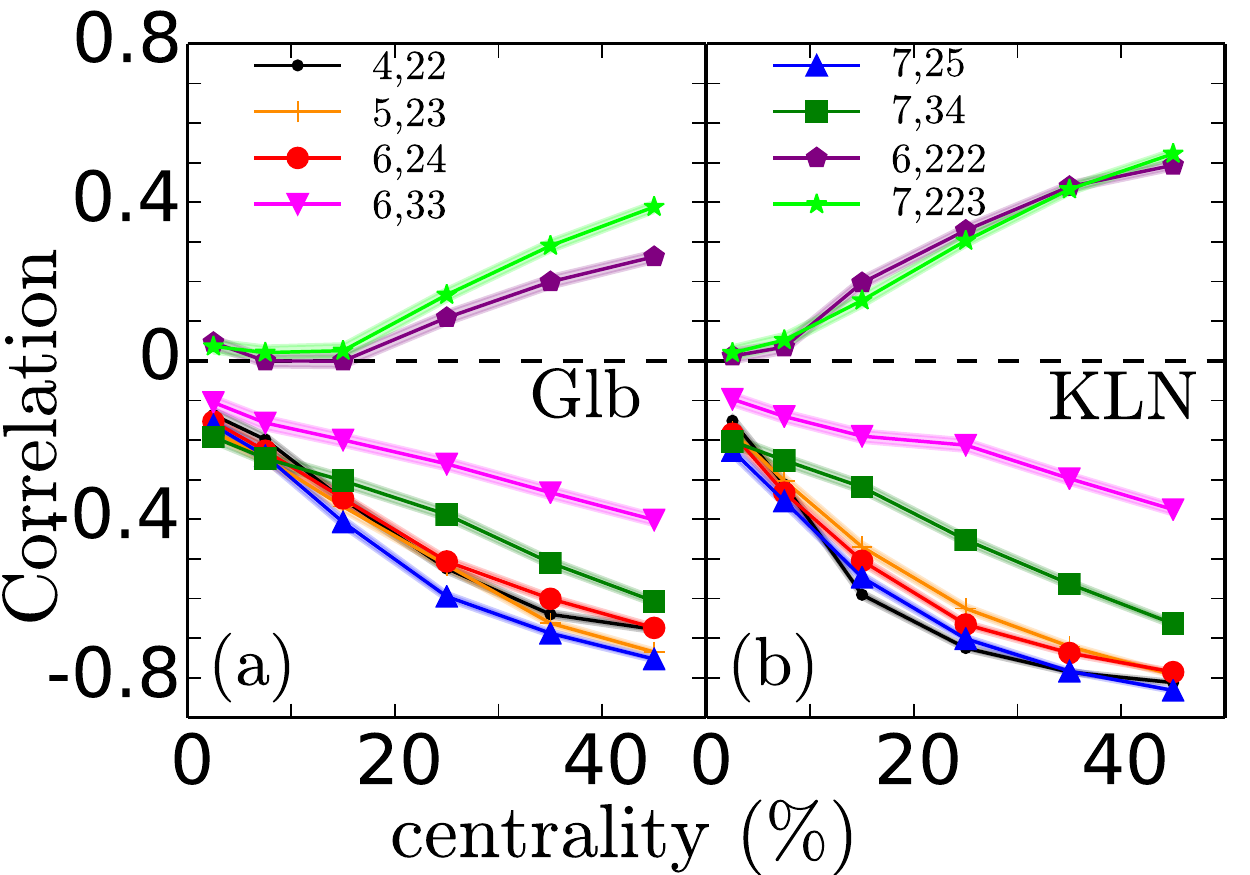}
    \caption{(Color online) Pearson correlation coefficients 
    $n,m_1\dots m_k\equiv\mathrm{Cor}(\mathcal{E}_n,\mathcal{E}_{m_1}\cdots\mathcal{E}_{m_k})$ 
    between the initial-state eccentricity coefficients $\mathcal{E}_n$ of harmonic order $n$ and bi- and 
    trilinear products of lower harmonic coefficients $m_i<n$, as functions of centrality. The correlation 
    coefficients are calculated for fluctuating initial density profiles obtained from (a) the MC-Glauber  
    and (b) the MC-KLN models, respectively.
    \label{F2}
}
\end{figure}

In Fig.~\ref{F2} we plot the Pearson correlation coefficients between these and a few other eccentricity coefficients of interest. The black line with small circles represents $\mathrm{Cor}(\mathcal{E}_4, \mathcal{E}_2^2)$; it shows a non-zero negative correlation that increases in magnitude with impact parameter. This non-negligible, even strong correlation (especially in non-central collisions) contradicts the above logical chain of arguments. This implies that the underlying assumption that in the decomposition (\ref{VlVnl_old}) the term $V_{nL}$ describes the linear response to $\mathcal{E}_n$ must be incorrect.
  
%
%

So if $V_{4L}$ is not the linear hydrodynamic response to $\mathcal{E}_4$, what is it? To further explore this question we recall that, since $V_{4L}$ and $V_2^2$ (and similarly $V_{5L}$ and $V_2V_3$) are statistically uncorrelated (as documented in Fig.~\ref{F1}), the rms values of their magnitudes can be computed from the relation \cite{Yan:2015jma}
\begin{equation}
\begin{split}
v_{4L}\{2\} &\equiv \sqrt{\langle v_{4L}^2 \rangle} = \sqrt{\langle v_4^2 \rangle - \frac{(\mathrm{Re} \langle V_4 V_2^{*2} \rangle)^2}{\langle v_2^4 \rangle}},\\
v_{5L}\{2\} &\equiv \sqrt{\langle v_{5L}^2 \rangle} = \sqrt{\langle v_5^2 \rangle - \frac{(\mathrm{Re} \langle V_5 V_2^* V_3^* \rangle)^2}{\langle v_2^2 v_3^2 \rangle}}.
\end{split}
\label{VnL}
\end{equation}
(These relations make the additional assumption $\langle V_{nL}\rangle\eq0$ which is natural if we take for granted that $V_{nL}$ carries a random phase factor $e^{in\Phi_\mathrm{RP}}$ from the direction of the impact parameter in the collision event.) We note in passing that taking the real part in the numerator of the second term under the square root is redundant since the corresponding imaginary part vanishes in the limit of large event samples by reflection symmetry of the underlying probability distribution with respect to the reaction plane.

\begin{figure}[!htb]
    \includegraphics[width=0.95\linewidth]{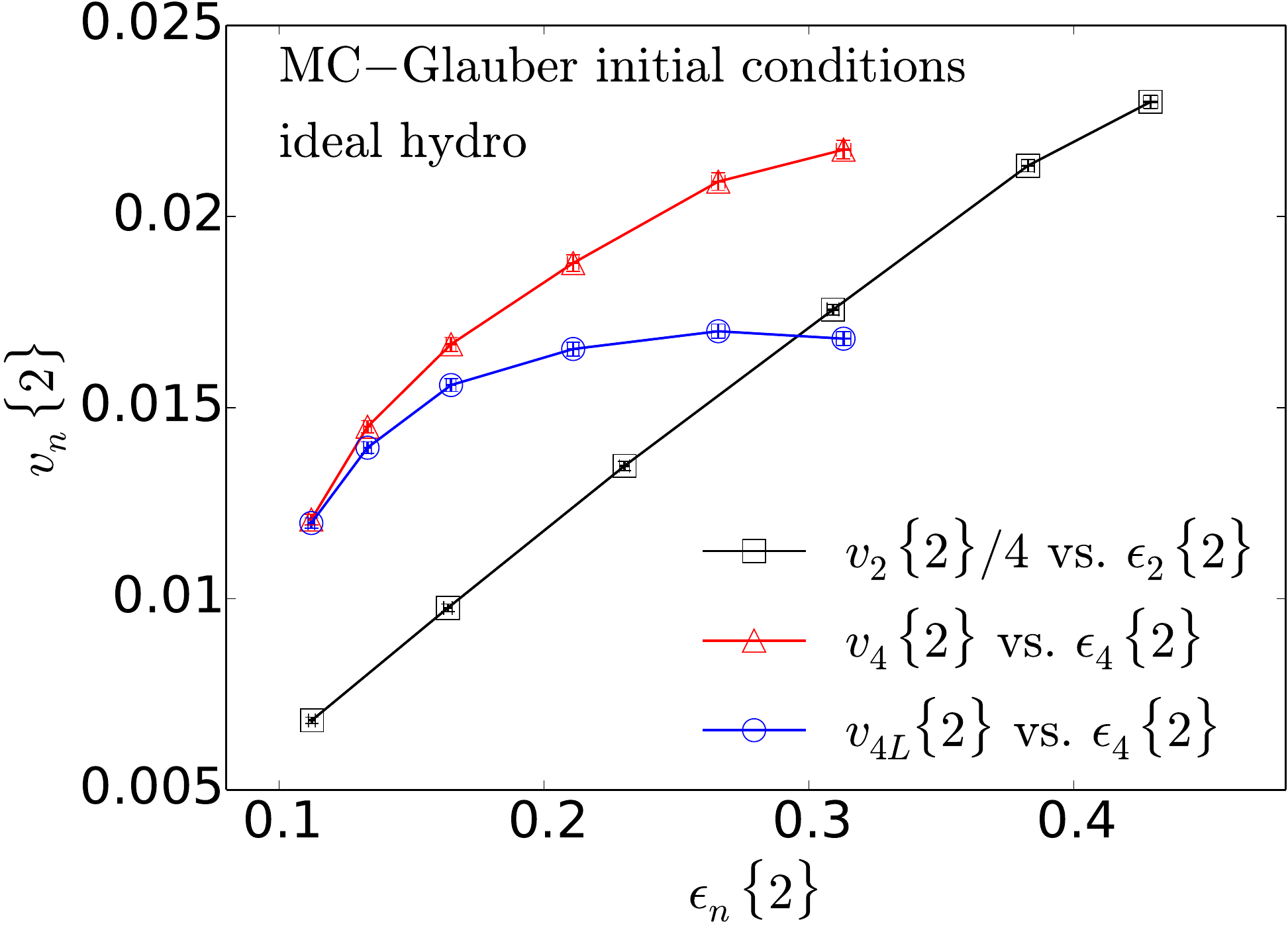}
    \caption{(Color online)
    $v_2\{2\}$, $v_4\{2\}$ and $v_{4L}\{2\}$ as functions of their corresponding eccentricities 
    $\epsilon_n\{2\}$. Each point corresponds to a fixed centrality bin (from left to right: 0-5\%, 
    5-10\%, 10-20\%, 20-30\%, 30-40\% and 40-50\%), each containing 2000 MC-Glauber events 
    evolved as ideal fluids. $v_2\{2\}$ has been divided by 4 to fit into the same plot. See text for 
    discussion.
    \label{F3}
}
\end{figure}

Using Eq.~(\ref{VnL}) we plot in Fig.~\ref{F3} $v_{4L}\{2\}$ together with the full second-order cumulant flows $v_4\{2\}$ and $v_2\{2\}$ as functions of their corresponding eccentricities, in order to check their linearity. The black circles demonstrate the well-known almost perfect linearity between the elliptic flow and the initial ellipticity. In contrast, the dependence of the full 4th-order flow $v_4\{2\}$ depends quite non-linearly on its corresponding eccentricity $\epsilon_4\{2\}$. This is also well-known and usually ascribed to increasingly important non-linear mode mixing with elliptic flow at larger impact parameters. What is surprising but supports the conclusion drawn above is that the so-called ``linear'' part $v_{4L}\{2\}$ exhibits even stronger nonlinearities than the full 4th-order flow when plotted as a function of $\epsilon_4\{2\}$. 

%
%

\begin{figure*}[t]
    \includegraphics[width=0.9\linewidth]{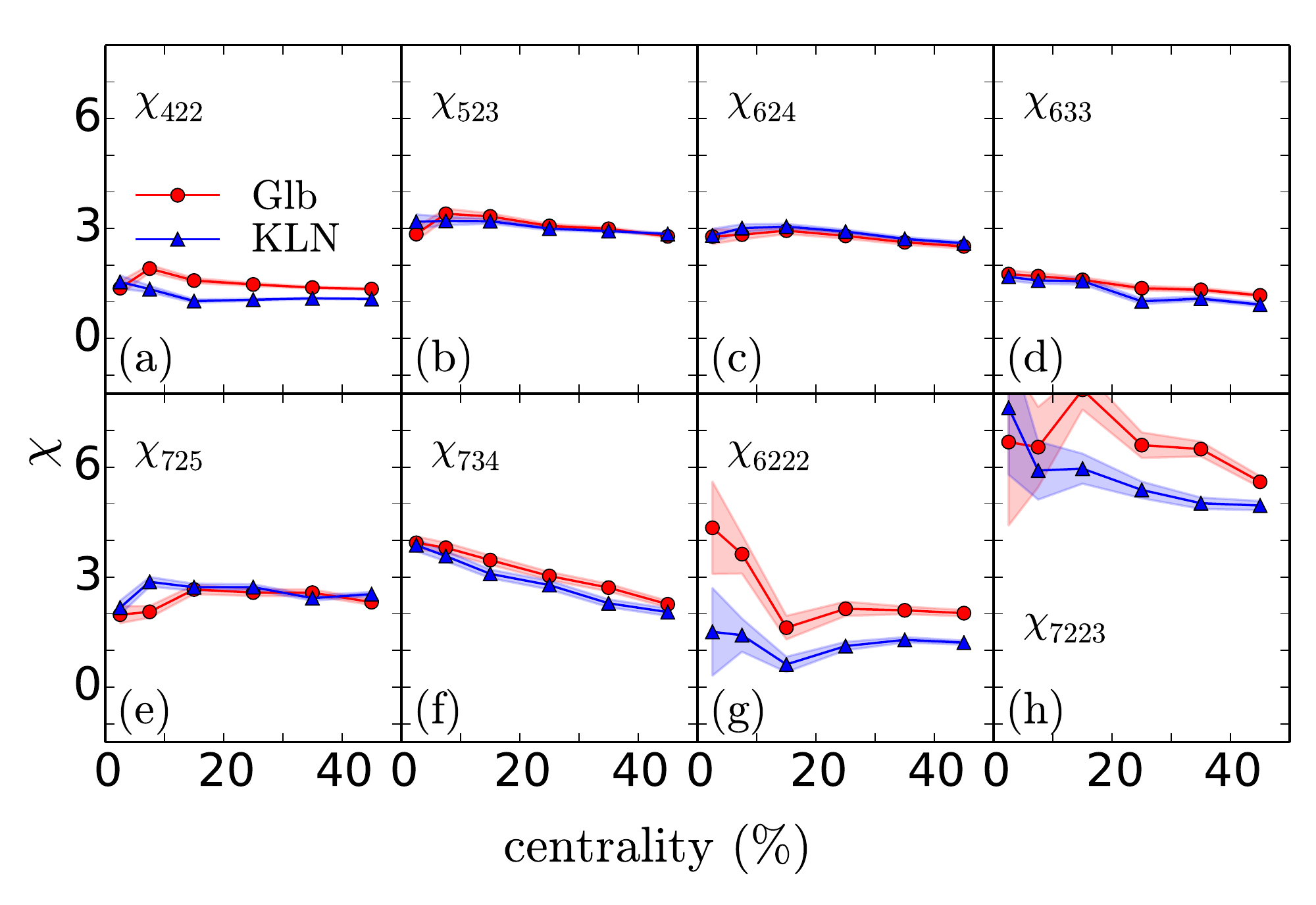}
    \caption{(Color online)
    The nonlinear response coefficients given in Eqs.~(\ref{chi_old}) and (\ref{eq13})-(\ref{eq18}) as 
    functions of centrality, for Pb-Pb collisions at $\sqrt{s} = 2.76\,A$\,TeV. Filled circles 
    (triangles) represent ideal fluid dynamical results using MC-Glauber (MC-KLN) initial conditions. 
    \label{F4}
}
\end{figure*}

From here on we will therefore consider the subscript $L$ on $V_{nL}$ to mean ``leading'' (in the decomposition (\ref{VlVnl_old})) or ``left-over'' rather than ``linear''. While this analysis does not provide a full answer to what drives the leading component $V_{nL}$, we reiterate the one aspect of this decomposition that matters for the rest of the paper: The leading (or left-over) terms $V_{nL}$ are statistically uncorrelated with the nonlinear mode coupling terms, and they average to zero. Following the arguments of \cite{Yan:2015jma} this allows to define and isolate the so-called non-linear mode coupling coefficients  
\begin{eqnarray}
\chi_{422} &=& \frac{\mathrm{Re} \langle V_4 (V_2^*)^2 \rangle}{\langle v_2^4 \rangle},\quad
\chi_{523} = \frac{\mathrm{Re} \langle V_5 V_2^* V_3^* \rangle}{\langle v_2^2 v_3^2 \rangle},
\nonumber\\
\chi_{633} &=& \frac{\mathrm{Re} \langle V_6 (V_3^*)^2 \rangle}{\langle v_3^4 \rangle},\quad
\chi_{6222} = \frac{\mathrm{Re} \langle V_6 (V_2^*)^3 \rangle}{\langle v_2^6 \rangle},
\nonumber\\
\chi_{7223} &=& \frac{\mathrm{Re} \langle V_7 (V_2^*)^2 V_3^* \rangle}{\langle v_2^4 v_3^2 \rangle}.
\label{chi_old}
\end{eqnarray}
Fig.~\ref{F4} displays these mode coupling coefficients (together with a few additional ones defined further below) as functions of centrality, using ideal fluid dynamics with MC-Glauber and MC-KLN initial conditions. This figure does not support the suggestion by Yan and Ollitrault \cite{Yan:2015jma} that, in general, they should be independent of the initial condition model: While $\chi_{422}$ (a), $\chi_{633}$ (d), and especially $\chi_{523}$ (b) indeed exhibit only weak sensitivity to the initial-state model, $\chi_{6222}$ (g) and $\chi_{7223}$ (h) differ significantly between MC-Glb and MC-KLN initial conditions. Coupling to a product of three $V_n$ vectors, these last two coefficients have larger statistical errors than the others which couple to only two other $V_n$ vectors, but their difference between the MC-Glb and MC-KLN initial conditions is clearly visible and statistically significant.
%
%

\section{Including additional mode coupling terms}
\label{sec3}
In the previous studies \cite{Teaney:2012ke,Yan:2015jma} the authors only considered those nonlinear couplings that involved contributions from the two largest flow harmonics $V_2$ and $V_3$ (which themselves are known to be dominated by linear response). This treatment seems incomplete. In this Section, we therefore add additional bilinear coupling terms of $V_n$ to lower-order flows $V_m$ ($m<n$) to the decomposition (\ref{VlVnl_old}): 
\begin{eqnarray}
\label{VlVnl_new}
V_n &=& V_{nL} + \sum_{k_1+k_2=n} \chi_{nk_1k_2} V_{k_1} V_{k_2} 
\nonumber\\
 && + \sum_{k_1+k_2+k_3=n} \chi_{nk_1k_2k_3} V_{k_1} V_{k_2} V_{k_3}.
\label{VnDecomposite}
\end{eqnarray}%
Trilinear couplings are still restricted to $k_i=2$ and 3. To test the importance of the various mode coupling terms in this decomposition we compute the Pearson correlation coefficients between $V_n$ and its possibly contributing mode coupling terms, analogous to the eccentricity correlation coefficients shown in Fig. \ref{F2}.
%
\begin{figure}[t]
    \includegraphics[width=\linewidth]{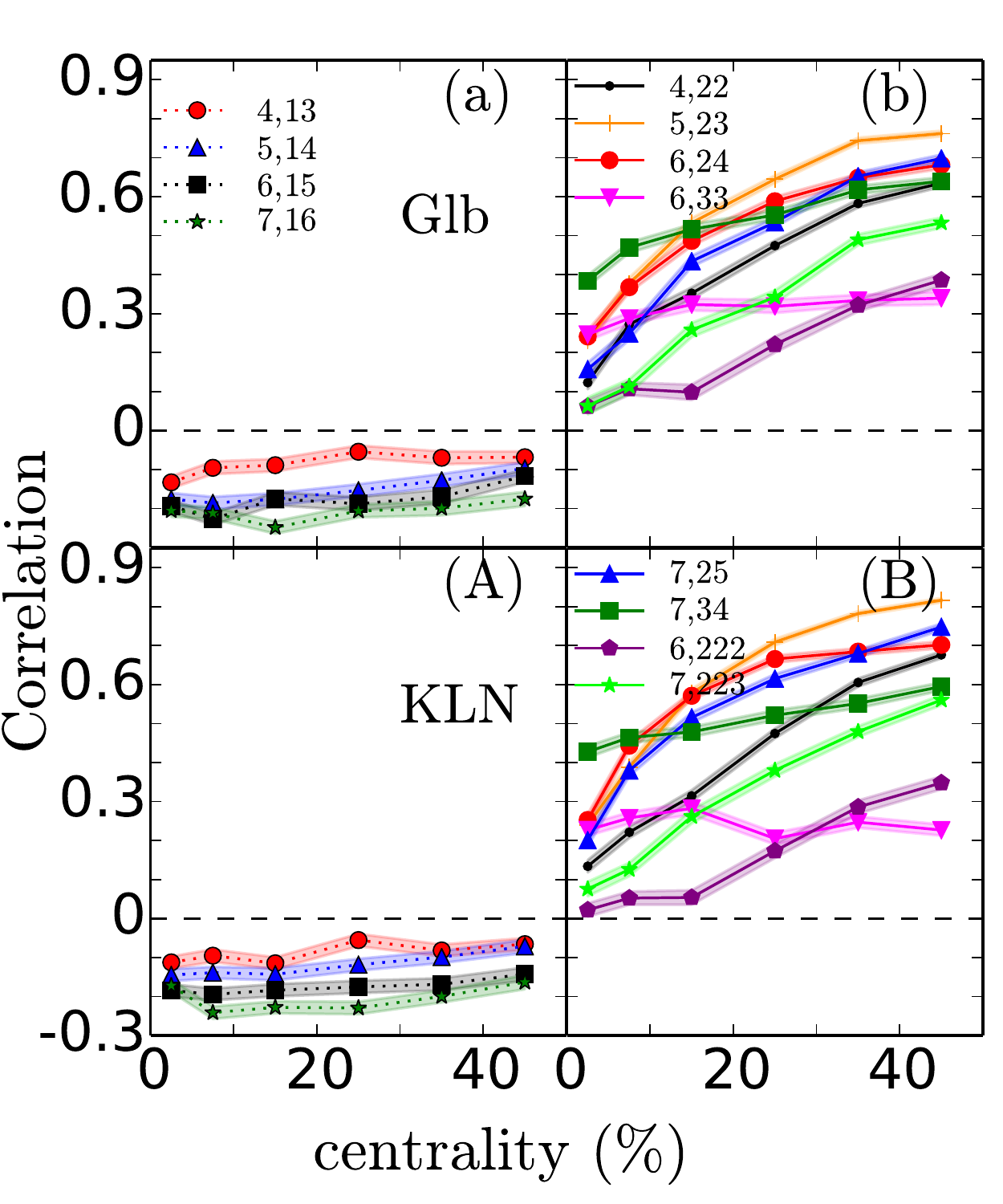}
    \caption{(Color online) Pearson correlation coefficients 
    $n,m_1\dots m_k\equiv\mathrm{Cor}(V_n, V_{m_1}\cdots V_{m_k})$ between the full $V_n$ and 
    their mode coupling contributions, as defined in Eq.~(\ref{complex_correlation_simple}), 
    as functions of centrality for Pb-Pb collisions at $\sqrt{s}{\,=\,}2.76\,A$\,TeV. The correlation coefficients 
    are obtained from the output of ideal fluid dynamical simulations with (a,b) MC-Glauber and (A,B) 
    MC-KLN initial conditions, respectively. 
    \label{F5}
}
\end{figure}
%

The correlation between $V_n$ and its possible mode coupling contributions  is shown in Fig.~\ref{F5} as a function of centrality, using ideal fluid dynamics with MC-Glauber and MC-KLN initial conditions for Pb-Pb collisions at 2.76\,$A$\,TeV. The left panels (a,A) show the correlation of $V_n$ with bilinear mode-coupling terms of the type $V_1 V_{n{-}1}$ where $V_1$ is the directed flow. Unlike higher harmonics, $V_1$ is affected by transverse momentum conservation. Considering units and normalization, we modify the weight used in \cite{Luzum:2010fb} to subtract the global momentum conservation contribution and calculate $V_1$ as follows \cite{Shen:2013cca}:
\begin{equation}
V_1 \equiv 
\frac{ \int p_T dp_T d \phi
        \left( \frac{p_T}{ \langle p_T \rangle} - \frac{\langle p_T^2 \rangle}{ \langle p_T \rangle^2}\right) 
        e^{ i \phi}\,\frac{dN}{dyp_Tdp_Td \phi}}
       {\int p_T dp_T d \phi 
       \left(\frac{p_T}{ \langle p_T \rangle} - \frac{ \langle p_T^2 \rangle }{ \langle p_T \rangle ^2}\right) 
       \frac{dN}{dyp_Tdp_Td \phi}}.
\label{V1def}
\end{equation}
Here angular brackets denote an average over particles in the desired $p_T$ range (over which $V_1$ is integrated) within a single event. Figs.~\ref{F5}a,A show that the correlations between $V_n$ and $V_1 V_{n{-}1}$ are all weak, for both MC-Glauber (a) and MC-KLN (A) initial conditions. Bilinear coupling terms involving the directed flow $V_1$ are therefore from now on ignored in Eq.~(\ref{VlVnl_new}).

Figs.~\ref{F5}b,B show that the new terms correlate equally strongly with $V_n$ as the original bilinear coupling terms in Eqs.~\ref{VlVnl_old}.  We will therefore include these terms in the following discussion of the mode coupling coefficients. In central collisions, none of the mode coupling terms are particularly strongly correlated with the full $V_n$, and in Sec.~\ref{sec2}, Fig.~\ref{F1} we found that they are also uncorrelated with the leading contributions on the right hand sides of the decompositions (\ref{VnDecomposite}). This strongly suggests that in central collisions (where according to Fig.~\ref{F2} eccentricity coefficients of different harmonic order are also seen to be essentially uncorrelated) $V_n$ is indeed dominated by linear response. The correlation between the total $V_n$ and its non-linear mode coupling contributions increases, however, with impact parameter. Since the leading term and the non-linear mode coupling terms are uncorrelated, this must mean that at large impact parameters $V_n$ is dominated by non-linear mode coupling. This is consistent with the conclusions of Ref.~\cite{Teaney:2012ke}.

We observe that all correlations of $V_n$ with mode coupling terms that couple to the elliptic flow $V_2$ increase with impact parameter. This reflects the concurrent growth of $V_2$, driven by the stronger elliptic deformation of the nuclear overlap zone in peripheral collisions. In contrast, the correlation of $V_6$ to the quadratic coupling of $V_3$ with itself is almost independent of collision centrality, again consistent with the much weaker centrality dependence of $V_3$ which is dominated by fluctuations rather than geometry \cite{Qiu:2011iv, Alver:2010gr}. Trilinear coupling terms exhibit generically smaller correlation coefficients with $V_n$ than bilinear terms, even if they involve the elliptic flow and thus grow together with $V_2$ in more peripheral collisions.

With the restrictions suggested by these observations, the decomposition Eq.~(\ref{VlVnl_new}) agrees with Eq.~(\ref{VlVnl_old}) for $V_4$ and $V_5$, while for $V_6$ and $V_7$ it generalizes to 
\begin{equation}
\begin{split}
  V_6 & = V_{6L} + \chi_{624} V_2 V_4 + \chi_{633} V_3^2 + \chi'_{6222}  V_2^3\\
         & = V_{6L} + \chi_{624} V_2 \bigl(V_{4L}{+}\chi_{422} V_2^2\bigr) 
                           + \chi_{633} V_3^2 + \chi'_{6222} V_2^3\\
         & \equiv V_{6L} + \chi_{624} V_2 V_{4L} + \chi_{633} V_3^2 + \chi_{6222}  V_2^3\,, \\
  V_7 & = V_{7L} + \chi_{725} V_2 V_5 + \chi_{734} V_3 V_4 + \chi'_{7223} V_2^2 V_3 \\
  	 & = V_{7L} + \chi_{725} V_2 \Bigl(V_{5L}{+}\chi_{523} V_2 V_3\bigr) \\
	 &\qquad\quad
	                    + \chi_{734} V_3 \bigl(V_{4L}{+}\chi_{422} V_2^2\bigr) 
	                    + \chi'_{7223} V_2^2 V_3 \\
	 & \equiv V_{7L} + \chi_{725} V_2 V_{5L} + \chi_{734} V_3 V_{4L} + \chi_{7223}  V_2^2 V_3\,.
\end{split}
\label{chi_n}
\end{equation}
The assumption that we can ignore correlations between the leading terms $V_{nL}$ and all mode coupling terms, together with $\langle V_{nL}\rangle\eq0$, leads to the following relations:
\begin{equation}
\begin{split}
  \langle V_6 V_2^* V_4^* \rangle 
  &= \chi_{624} \langle v_2^2 \rangle \langle v_{4L}^2 \rangle 
     + \chi_{6222} \chi_{422} \langle v_2^6 \rangle, \\
  \langle V_6 V_3^{*2} \rangle 
  &= \chi_{633} \langle v_3^4 \rangle, \\
  \langle V_6 V_2^{*3} \rangle 
  &= \chi_{6222} \langle v_2^6 \rangle, \\
  \langle V_7 V_2^* V_5^* \rangle 
  &= \chi_{725} \langle v_2^2 \rangle \langle v_{5L}^2 \rangle 
     + \chi_{7223} \chi_{523} \langle v_2^4 v_3^2 \rangle, \\
  \langle V_7 V_3^* V_4^* \rangle 
  &= \chi_{734} \langle v_3^2 \rangle \langle v_{4L}^2 \rangle 
     + \chi_{7223} \chi_{422} \langle v_2^4 v_3^2 \rangle, \\
  \langle V_7 V_2^{*2} V_3^* \rangle 
  &= \chi_{7223} \langle v_2^4 v_3^2 \rangle.
\end{split}
\label{before_coefficients}
\end{equation}
The validity of these relations is checked and found to hold in Appendix~\ref{appa}. Using the definitions (\ref{VnL}) of $\langle v_{4L}^2\rangle$ and $\langle v_{5L}^2 \rangle$ the mode coupling coefficients $\chi$ can be isolated from Eqs.~(\ref{before_coefficients}) as follows:
\begin{eqnarray}
\label{eq13}
  &&\chi_{624} =
        \mathrm{Re}\frac{\langle V_6V_2^*V_4^* \rangle \langle v_2^4 \rangle 
                                  - \langle V_6 V_2^{*3} \rangle \langle V_4 V_2^{*2} \rangle}
                                   {\bigl( \langle v_4^2 \rangle \langle v_2^4 \rangle%
                                  {-}\langle V_4 V_2^{*2} \rangle^2\bigr)\,\langle v_2^2 \rangle},
\\
\label{eq14}
  &&\chi_{633} = \frac{\mathrm{Re} \langle V_6 V_3^{*2} \rangle}{\langle v_3^4 \rangle},
\\
\label{eq15}
  &&\chi_{6222} = \frac{\mathrm{Re} \langle V_6 V_2^{*3} \rangle}{\langle v_2^6 \rangle},
\\
\label{eq16}
  &&\chi_{725} = 
        \mathrm{Re}\frac{\langle V_7 V_2^* V_5^* \rangle \langle v_2^2 v_3^2 \rangle 
                                   - \langle V_7 V_2^{*2} V_3^*\rangle \langle V_5 V_2^* V_3^* \rangle}
                                   {\bigl(\langle v_5^2 \rangle \langle v_2^2 v_3^2 \rangle%
                                     {-}\langle V_5 V_2^* V_3^* \rangle^2\bigr)\,\langle v_2^2 \rangle},\ \ \quad
\\
\label{eq17}
  &&\chi_{734} = 
        \mathrm{Re}\frac{\langle V_7 V_3^* V_4^* \rangle \langle v_2^4 \rangle 
                                   - \langle V_7 V_2^{*2} V_3^* \rangle \langle V_4 V_2^{*2} \rangle}
                                   {\bigl(\langle v_4^2 \rangle \langle v_2^4 \rangle%
                                   {-}\langle V_4 V_2^{*2}\rangle ^2\bigr)\,\langle v_3^2 \rangle},
\\
\label{eq18}
  &&\chi_{7223} = \frac{\mathrm{Re} \langle V_7 V_2^{*2} V_3^* \rangle}{\langle v_2^4 v_3^2 \rangle}.
\end{eqnarray}
%
%
\begin{figure*}[!htb]
    \includegraphics[width=0.8\linewidth]{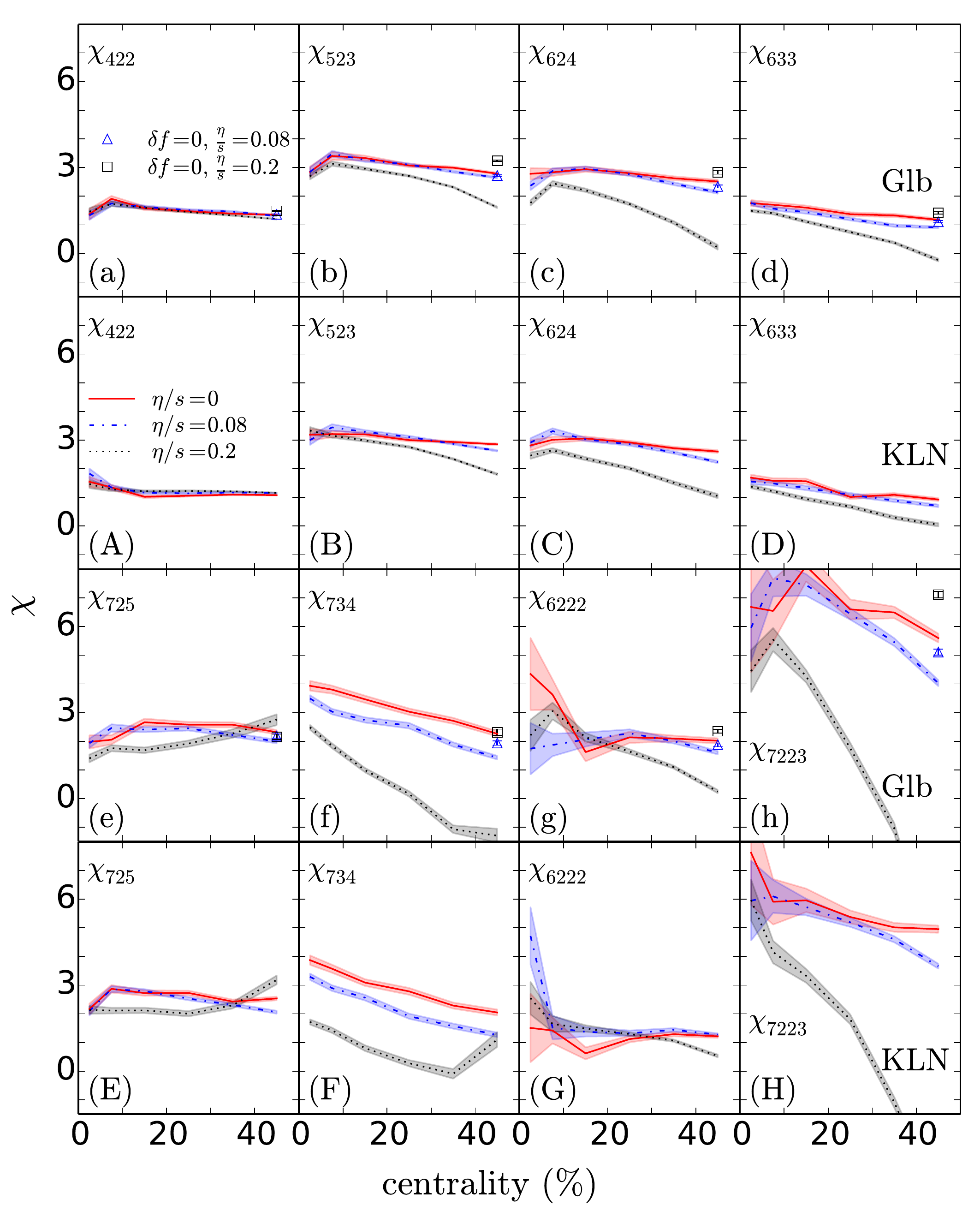}
    \caption{(Color online)
      Nonlinear response coefficients defined by Eqs.~(\ref{chi_old}), (\ref{eq13}), (\ref{eq16}) and 
      (\ref{eq17}). Solid red lines correspond to ideal hydrodynamics while dash-dotted blue and 
      dotted black lines correspond to viscous hydrodynamics with $\eta / s = 0.08$ and 0.2, 
      respectively. Panels labeled by lower case (upper case) letters show results from MC-Glb 
      (MC-KLN) initial conditions, respectively. In the MC-Glb panels, the open symbols indicate 
      the values of the mode coupling coefficients in the 40-50\% centrality bin if the viscous 
      correction $\delta f$ at freeze-out is ignored (see text); blue triangles and black squares 
      correspond to $\eta / s = 0.08$ and 0.2, respectively.        
    \label{F6}
}
\end{figure*}%
%
The expressions for $\chi_{633}$, $\chi_{6222}$ and $\chi_{7223}$ agree with those derived before in Eqs.~(\ref{chi_old}). The additional mode coupling coefficients $\chi_{624}$, $\chi_{725}$ and $\chi_{734}$, which supplement those listed in Eqs.~(\ref{chi_old}), are shown in panels c, e, f of Fig.~\ref{F4}. Like the other bilinear mode coupling terms in that figure, they are very similar for the MC-Glauber and MC-KLN initial-state models, although not completely independent of the initial conditions. Stronger sensitivity to the initial fluctuation spectrum is clearly observed in the trilinear coupling terms (panels g and h), as already noted. 

In Fig.\,\ref{F6} we explore the sensitivity of the mode-coup\-ling coefficients Eqs.~(\ref{chi_old}), (\ref{eq13}), (\ref{eq16}) and (\ref{eq17}) to the specific shear viscosity $\eta/s$ of the evolving hydrodynamic medium. We performed this study for both MC-Glb (panels labeled by lower case letters) and MC-KLN initial conditions (panels labeled by upper case letters), and display in the figure the corresponding results in pairs of panels arranged directly above each other. Results are shown for three different choices of $\eta/s$: $\eta/s\eq0$ (ideal fluid, red solid lines), $\eta/s\eq0.08$ (``minimal'' specific shear viscosity, blue dash-dotted lines), and $\eta/s\eq0.2$ (black dotted lines). Generically, the nonlinear mode-mixing coefficients decrease with increasing shear viscosity, as had already been observed by Yan and Ollitrault in Fig.~2 of Ref.~\cite{Yan:2015jma}. However, we find a weaker sensitivity to $\eta/s$, and also disagree on (some of) their magnitudes.\footnote{For example, we find $\chi_{633}\lesssim\chi_{6222}$ whereas Yan and Ollitrault \cite{Yan:2015jma} find the opposite.} For 0-5\% centrality our results for the trilinear coupling coefficients fluctuate a lot and are statistically quite uncertain; much higher event statistics would be needed to significantly improve this situation. The differences between our results and those from Ref.~\cite{Yan:2015jma} may indicate shortcomings of the approach used by Yan and Ollitrault who, instead of evolving genuinely bumpy initial conditions obtained from a Monte Carlo sampling of the initial nucleon positions in the colliding nuclei, use smooth initial Gaussian density profiles that are azimuthally deformed ``in order to produce anisotropic flow in the desired harmonic'' \cite{Yan:2015jma}. 

We emphasize the complete insensitivity to shear viscosity of the bilinear coupling $\chi_{422}$ of elliptic flow $V_2$ coupling to itself to produce quadrangular flow $V_4$: As shown in panels a, A of Fig.~\ref{F6}, the differences between its values for different shear viscosities are much smaller even than the differences between MC-Glb and MC-KLN initial conditions. This points to a hydrodynamic flow profile whose quadrangular deformation is very small (even in the ideal fluid case, without viscous damping) such that the $V_4$ of the finally observed hadron momentum distribution is entirely dominated by the contribution generated at freeze-out via an elliptic deformation of the fluid velocity profile, as discussed in \cite{Borghini:2005kd}. 

The same does not hold for elliptic flow $V_2$ coupling with itself to produce $V_6$ ($\chi_{6222}$, shown in panels f, F), nor for the frequency-doubling mode-coupling of $V_3$ to itself to produce $V_6$ ($\chi_{633}$, shown in panels d, D). In fact, all other mode-coupling coefficients show significant sensitivity to $\eta/s$ (especially in non-central collisions). The question arises whether this sensitivity $\eta/s$ reflects shear viscous effects on the buildup of hydrodynamic flow during the entire evolution, or whether it is simply due to the viscous deviation $\delta f$ of the local phase-space distribution at freeze-out caused by the non-vanishing shear stress on the decoupling surface (which is also proportional to $\eta/s$). 

To help answer this question, we show in Figs.~\ref{F6}a-h the nonlinear mode-coupling coefficients calculated without the $\delta f$ contribution, for the 40-50\% centrality bin with MC-Glauber initial conditions. Comparing the open blue triangles (black squares) for $\delta f=0$ with the values of the blue dash-dotted (black dotted) lines (which include the $\delta f$ contribution) in the same centrality bin, we observe (in agreement with Refs.~\cite{Teaney:2012ke,Yan:2015jma}) large to very large $\delta f$ correction effects for all mode-coupling terms except the bilinear self coupling of $V_2$, $\chi_{422}$. In fact, by setting $\delta f$ to zero and thereby focusing on shear viscous effects on the hydrodynamic flow alone, we see that non-zero shear viscosity slightly increases the strength of mode-coupling effects (as observed before in \cite{Qiu:2012uy}), presumably by damping the hydrodynamic effects of event-by-event initial-state fluctuations that tend to decorrelate the event planes \cite{Qiu:2012uy}. This increase is more than compensated for by a large negative $\delta f$ contribution to the nonlinear coupling coefficients, which completely dominates the net sensitivity of these coefficients to $\eta/s$. The nonlinear coupling coefficients are therefore mostly sensitive to shear viscous stresses at freeze-out and less so to the value of the specific shear viscosity during the earlier evolution stages. This clearly limits their value as signatures of the transport properties of the evolving medium, irrespective of whether or not they depend on the spectrum of initial-state fluctuations.

\section{``Mode coupling'' between the initial eccentricities}
\label{sec4}
Anisotropic flow is the hydrodynamic response to the anisotropic and bumpy initial density profile. In this section we elucidate further to what extent the mode coupling effects represented by the nonlinear coupling coefficients defined in Eqs.~(\ref{chi_old},\ref{eq13}-\ref{eq18}) and studied in the preceding section are due to nonlinearities in this hydrodynamic response or already pre-exist among the eccentricity coefficients $\mathcal{E}_n$ of the initial fluctuating density distributions, due to geometric constraints on the initial fluctuation spectrum.    

%
\begin{figure*}[!htb]
    \includegraphics[width=0.8\linewidth]{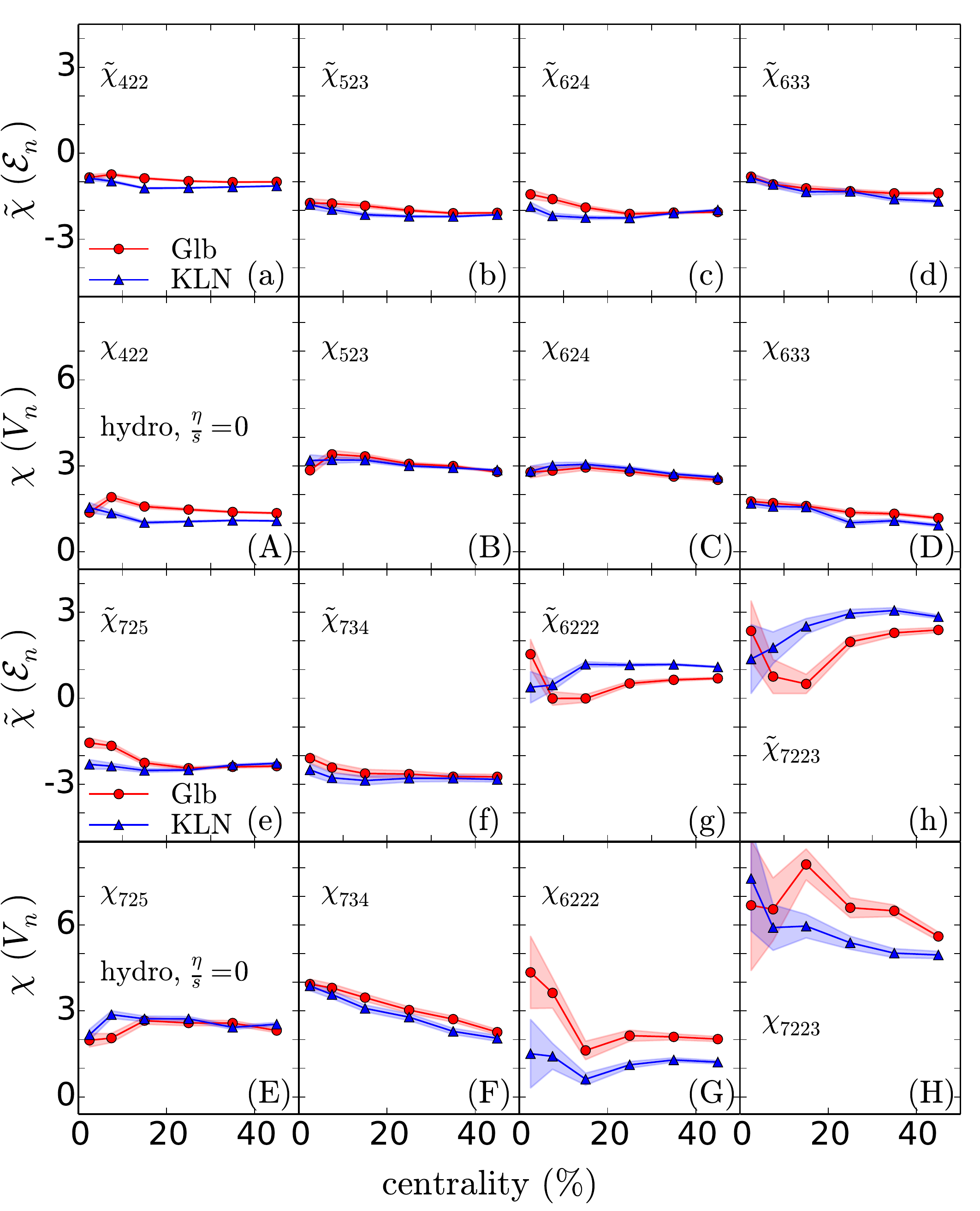}
    \caption{(Color online)
    Comparison of the centrality dependences of the nonlinear initial eccentricity ($\tilde\chi$, panels a-h)
    and final flow coupling coefficients from ideal fluid dynamics ($\chi$, panels A-H),
    with MC-Glb (circles) and MC-KLN (triangles) initial conditions, for Pb-Pb collisions at 
    $\sqrt{s}{\,=\,}2.76\,A$\,TeV. 
    \label{F7}
}
\end{figure*}
%

To this end we compare the Pearson correlation coefficients between higher-order anisotropic flows $V_n$ and their nonlinear contributions shown in Figs.~\ref{F5}b,B with those for the corresponding initial-state eccentricity coefficients $\mathcal{E}_n$ shown in Fig.~\ref{F2}a,b. One notices that the correlations between $\mathcal{E}_n$ and their contributions from bilinear coupling terms are all negative while the corresponding flow correlations are all positive. We will return to this observation further below. Although there are some quantitative differences between the two initial-state models, their qualitative features (in particular their centrality dependences) are very similar. Most importantly, the correlation coefficients between the initial-state eccentricities shown in Figs.~\ref{F2}a,b exhibit (except for their sign) a close similarity, in both magnitude and centrality dependence, with the corresponding correlation coefficients for the final anisotropic flows shown in Figs.~\ref{F5}b,B. This supports the conclusion above that the nonlinear flow-coupling coefficients are not dominated by nonlinear hydrodynamic response but, at least to a large part, ``pre-formed'' by geometric correlations among the initial eccentricity coefficients and linearly propagated into the final state.  

We follow this theme further and define the nonlinear eccentricity-coupling coefficients $\tilde{\chi}$ in analogy to the nonlinear flow-coupling coefficients Eqs.~(\ref{chi_old},\ref{eq13}-\ref{eq18}):
\begin{eqnarray}
\tilde{\chi}_{422} & =& \frac{\mathrm{Re} \langle \mathcal{E}_4 \mathcal{E}_2^{*2} \rangle}
                                           {\langle \epsilon_2^4 \rangle},\quad
\tilde{\chi}_{523} = \frac{\mathrm{Re} \langle \mathcal{E}_5 \mathcal{E}_2^* \mathcal{E}_3^* \rangle}
                                     {\langle \epsilon_2^2 \epsilon_3^2 \rangle},
\nonumber\\
\tilde{\chi}_{624} &=& 
     \mathrm{Re}\frac{\langle \mathcal{E}_6\mathcal{E}_2^*\mathcal{E}_4^* \rangle 
                                  \langle \epsilon_2^4 \rangle 
                                - \langle \mathcal{E}_6 \mathcal{E}_2^{*3} \rangle 
                                  \langle \mathcal{E}_4 \mathcal{E}_2^{*2} \rangle}
                                {\bigl(\langle\epsilon_4^2\rangle \langle\epsilon_2^4\rangle%
                                {-}\langle\mathcal{E}_4\mathcal{E}_2^{*2}\rangle^2\bigr)\,\langle \epsilon_2^2\rangle},
\nonumber\\\nonumber
\tilde{\chi}_{633} &=& 
     \frac{\mathrm{Re} \langle\mathcal{E}_6\mathcal{E}_3^{*2}\rangle}{\langle \epsilon_3^4 \rangle},
\end{eqnarray}
\begin{eqnarray}
\tilde{\chi}_{725} &=&
      \mathrm{Re}\frac{\langle\mathcal{E}_7\mathcal{E}_2^*\mathcal{E}_5^*\rangle 
                                   \langle\epsilon_2^2\epsilon_3^2\rangle 
                                   - \langle\mathcal{E}_7\mathcal{E}_2^{*2}\mathcal{E}_3^*\rangle 
                                   \langle\mathcal{E}_5\mathcal{E}_2^*\mathcal{E}_3^*\rangle}
                                 {\bigl(\langle\epsilon_5^2\rangle \langle\epsilon_2^2\epsilon_3^2\rangle%
                                  {-}\langle\mathcal{E}_5\mathcal{E}_2^*\mathcal{E}_3^*\rangle^2\bigr)\,
                                  \langle\epsilon_2^2\rangle},
\nonumber\\
\tilde{\chi}_{734} &=& 
    \mathrm{Re}\frac{\langle\mathcal{E}_7\mathcal{E}_3^*\mathcal{E}_4^*\rangle 
                                \langle\epsilon_2^4\rangle 
                                - \langle\mathcal{E}_7\mathcal{E}_2^{*2}\mathcal{E}_3^*\rangle 
                                  \langle\mathcal{E}_4\mathcal{E}_2^{*2}\rangle}
                               {\bigl(\langle\epsilon_4^2\rangle \langle\epsilon_2^4 \rangle%
                                {-}\langle\mathcal{E}_4\mathcal{E}_2^{*2}\rangle^2\bigr)\,\langle\epsilon_3^2\rangle},
\nonumber\\
\tilde{\chi}_{6222} &=& \frac{\mathrm{Re} \langle\mathcal{E}_6\mathcal{E}_2^{*3}\rangle}
                                            {\langle\epsilon_2^6\rangle},\quad
\tilde{\chi}_{7223} = \frac{\mathrm{Re} \langle\mathcal{E}_7\mathcal{E}_2^{*2}\mathcal{E}_3^*\rangle}
                                       {\langle\epsilon_2^4\epsilon_3^2 \rangle}.
\label{chi_ecc}
\end{eqnarray}

To eliminate the contribution to the final flow coefficients from the $\delta f$ correction at freeze-out (see discussion of Fig.~\ref{F6}), we compare in Fig.~\ref{F7} these nonlinear eccentricity coupling coefficients with the nonlinear flow coupling coefficients from {\em ideal} fluid dynamics, for both MC-Glauber and MC-KLN initial conditions. None of the coefficients exhibit strong centrality dependence. Differences between the final flow coupling coefficients $\chi$ from the two different initial state models appear to be mostly caused by analogous differences between the corresponding eccentricity coupling coefficients $\tilde\chi$ existing already in the initial state. As observed in Fig.~\ref{F4}, these differences are small for bilinear coupling coefficients but appear to be larger for trilinear couplings. 

The most important feature of Fig.~\ref{F7} is the sign change between the consistently negative values for $\tilde\chi$ and the positive values of $\chi$ for bilinear couplings. As already mentioned, this sign change between eccentricity and flow correlations is also seen in the Pearson correlation coefficients shown in Figs.~\ref{F2} and \ref{F5}. This observation is consistent with Refs.~\cite{Qiu:2012uy,Bhalerao:2013ina,Teaney:2013dta,Qiu:2013wca} where it was found that hydrodynamic evolution changes the sign of the correlations between the initial participant planes and the final flow planes that are associated with these coupling coefficients.%
\footnote{For example, $\langle\mathcal{E}_4\mathcal{E}_2^{*2}\rangle$ is an eccentricity-weighted
              average of the participant-plane correlator $\cos\bigl(4(\Phi_4{-}\Phi_2)\bigr)$.}
In Fig.~4.5 of Zhi Qiu's Ph.D. thesis \cite{Qiu:2013wca} it was shown that this sign change is genuinely related with a sign change between the final hydrodynamic flow plane and initial participant plane correlators, and not qualitatively changed by mode-mixing effects inherent in the Cooper-Frye formula for computing the momentum distributions and their anisotropies at freeze-out \cite{Borghini:2005kd}. This led the authors of Ref.~\cite{Qiu:2012uy} to conclude that this sign change is a robust signature of {\em nonlinear} hydrodynamic response to the initial density and associated pressure gradients. The discussion of Fig.~\ref{F6} above suggests that this nonlinear hydrodynamic response exhibits no strong sensitivity to the specific shear-viscosity of the hydrodnamically evolving medium, as almost all of the dependence of the nonlinear coupling coefficients between the final anisotropic flows on $\eta/s$ arises from the $\delta f$ correction at freeze-out (which depends on $\eta/s$ at freeze-out, not at earlier times).   

\section{Summary and further discussion}
\label{sec6}
We have presented a systematic hydrodynamic study of nonlinear mode coupling contributions to higher order anisotropic flows in 2.76\,$A$\,TeV Pb-Pb collisions at the LHC. We compared the relevant mode coupling coefficients between the final anisotropic flow vectors $V_n$ with the corresponding nonlinear coupling coefficients between the initial eccentricity vectors $\mathcal{E}_n$ which embody geometric correlations between different harmonic components of the fluctuations in the initial state. While the authors of \cite{Bhalerao:2014xra,Yan:2015jma} expected the mode coupling coefficients to be independent of initial conditions, we found that several of them exhibit non-negligible dependence on the model used to generate the fluctuating initial state. We also showed that qualitatively similar model dependence is already seen in the corresponding initial nonlinear eccentricity coupling coefficients, likely driven by somewhat different geometric constraints on the eccentricity fluctuation spectrum in the two initial-state models studied here (the Monte Carlo Glauber and KLN models) whose ensemble-averaged density profiles are known to differ. 

The calculations demonstrate significant dynamical evolution of the initial nonlinear eccentricity couplings to the final nonlinear flow couplings. In particular, all bilinear coupling coefficients coupling two lower-order harmonic coefficients to a higher-order one flip sign between the initial and final state. The initial nonlinear eccentricity coupling coefficients are closely related to participant-plane correlations in the initial state while the final nonlinear flow coupling coefficients reflect flow-plane correlations in the final state. Therefore, the sign change between the initial eccentricity and final flow coupling coefficients observed here confirms a similar sign change between the participant and flow plane correlations observed earlier in Refs.~\cite{Qiu:2012uy, Bhalerao:2013ina, Teaney:2013dta} and attributed to nonlinear hydrodynamic response.

%
\begin{figure*}[!htb]
    \includegraphics[width=\linewidth]{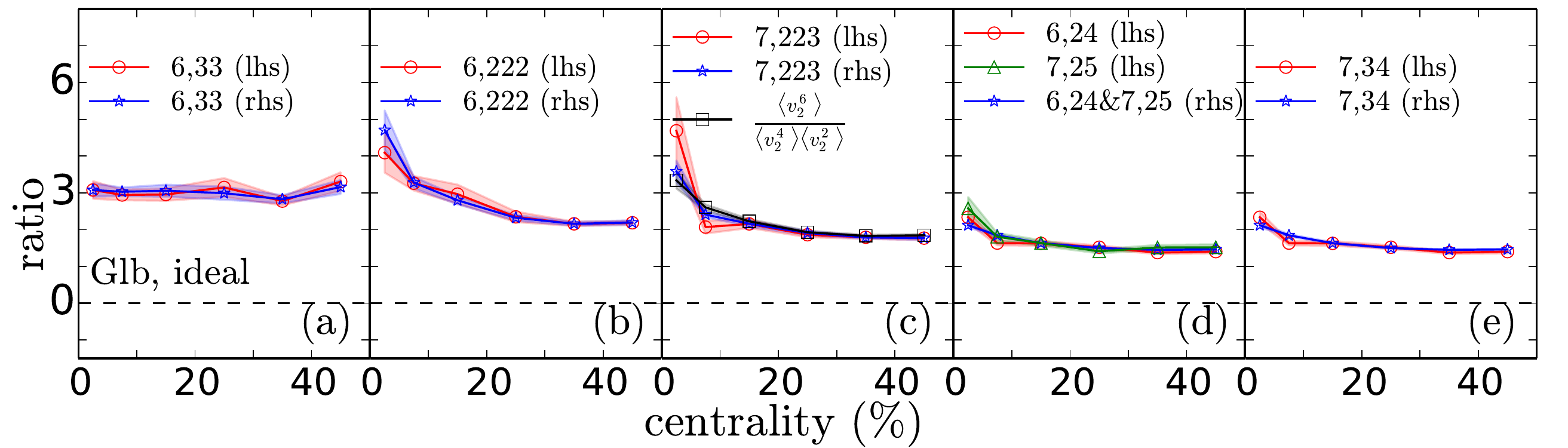}
    \vspace*{-5mm}
    \caption{(Color online)
    Test of Eqs.~(\ref{A1})-(\ref{A3}) and (\ref{A7})-(\ref{A9}), using results from ideal hydrodydynamic 
    simulations for Pb-Pb collisions at $\sqrt{s}\eq2.76\,A$\,TeV with MC-Glauber initial conditions. 
    Circles (stars) denote the left-hand (right-hand) sides of (a) Eq.~(\ref{A1}), (b) Eq.~(\ref{A2}), 
    (c) Eq.~(\ref{A3}), (d) Eqs.~(\ref{A7},\ref{A8}) (which have the same right-hand sides), and 
    (e) Eq.~(\ref{A9}). The squares in panel (c) show the alternate version of the r.h.s. corresponding 
    to the second equality sign in Eq.~(\ref{A3}). 
    \label{F8}
}
\end{figure*}
%

A study of the sensitivity of the nonlinear mode coupling coefficients between the final anisotropic flow vectors $V_n$ showed very weak dependence on the shear viscosity of the evolving medium but very strong sensitivity, especially in non-central collisions, to the shear viscosity at freeze-out, through the viscous correction $\delta f$ to the local distribution function on the freeze-out surface which depends on the shear stress on this surface. Related observations made in Refs.~\cite{Teaney:2012ke,Yan:2015jma} support these findings. This eliminates these nonlinear mode coupling coefficients from the list of candidates for observables that might provide insights on the transport properties of the evolving medium independent of the model used to simulate the (not very well known and not directly measurable) initial fluctuation spectrum. 

\begin{acknowledgments}
We thank Christopher Plumberg, Chun Shen and Hong Zhang for fruitful discussions and valuable comments. Jean-Yves Ollitrault and Li Yan are gratefully acknowledged for pointing out an error in the original manuscript related to the definition of the additional mode coupling coefficients in Eqs.~(\ref{chi_n}). This work was supported by the Department of Energy, Office of Science, Office of Nuclear Physics under Award No. \rm{DE-SC0004286}; computing resources were generously provided by the Ohio Supercomputer Center \cite{OhioSupercomputerCenter1987}. J.Q. acknowledges support by Harbin Institute of Technology through its PhD Short-term Academic Visiting Program. 
\end{acknowledgments}


\appendix
\vspace*{-3mm}
\section{Absence of correlations between $V_{nL}$ and mode coupling terms}
\label{appa}
\vspace*{-3mm}

\begin{figure*}[!htb]
    \includegraphics[width=0.88\linewidth]{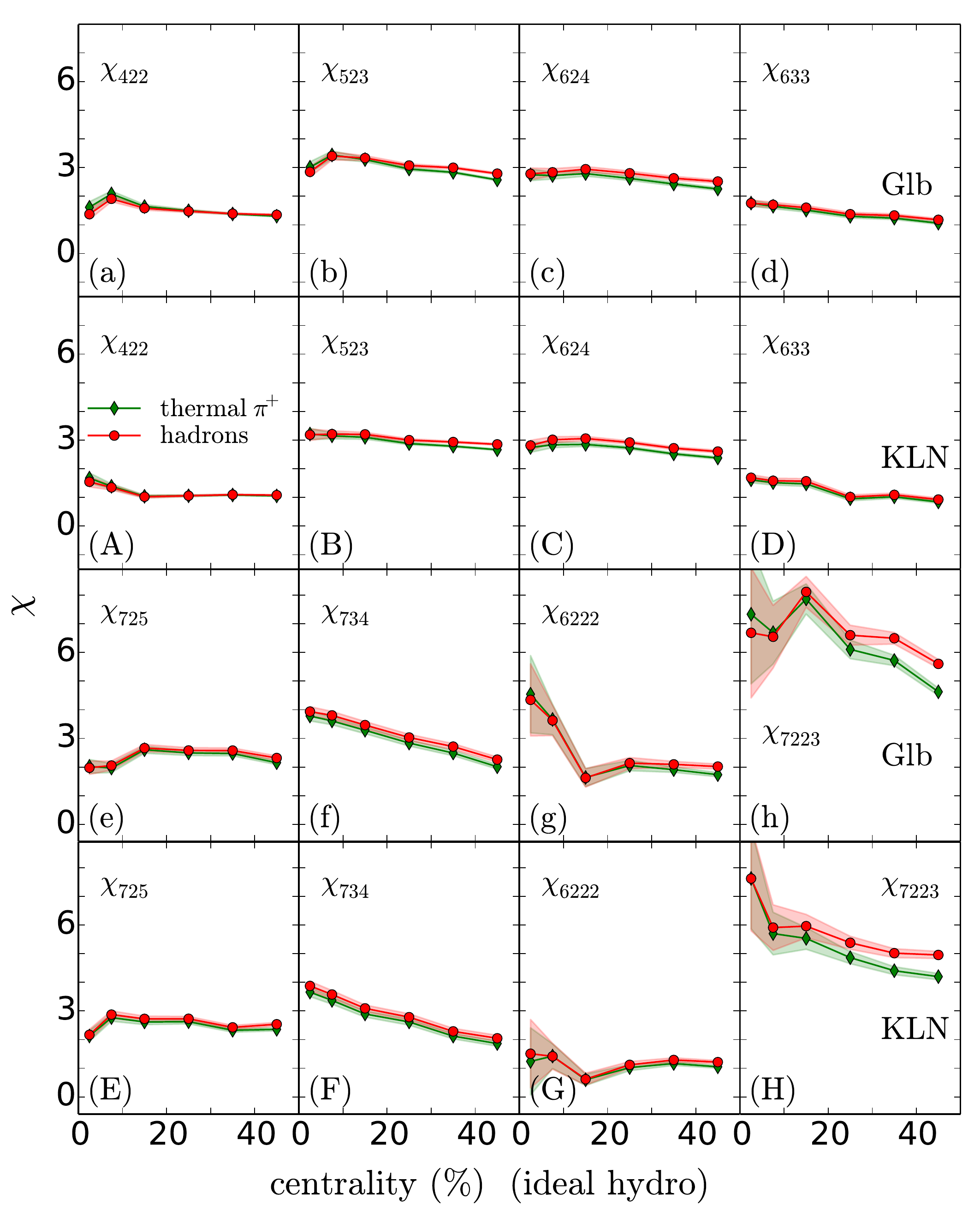}
    \caption{(Color online)
    Mode coupling coefficients from ideal hydrodynamic simulations for Pb-Pb collisions at 
    2.76\,$A$\,TeV with MC-Glauber (panels a-h) and MC-KLN (panels A-H) initial conditions, 
    computed from flow vectors $V_n$ for directly emitted (``thermal'') $\pi^+$ (green diamonds) 
    and for all charged hadrons (red circles). 
     \label{F9}
}
\end{figure*}

In this appendix we provide a numerical proof that in Eqs.~(\ref{chi_n}) the leading terms $V_{nL}$ are uncorrelated with all of the mode coupling terms. To this end, we construct ratios of moments of the type $\frac{\langle V_n\,T\,A \rangle}{\langle V_n\,T \rangle \langle A \rangle}$ where $n$ labels the harmonic order of the term $V_{nL}$ in question, $T$ stands for the mode coupling product whose statistical independence of $V_{nL}$ we want to test, and $A$ denotes auxiliary factors involving only the magnitudes $v_m$ of flow coefficients. We also use the known statistical independence of the flow angles between the elliptic and triangular flow vectors $V_2$ and $V_3$. Next, we decompose $V_n$ according to Eqs.~(\ref{VlVnl_old}) for $V_4$ and $V_5$ and according to Eqs.~(\ref{chi_n}) for $V_6$ and $V_7$. Using the fact (established in Fig.~\ref{F1}) that $V_{4L}$ and $V_{5L}$ are uncorrelated with $V_2^2$ and $V_2V_3$, respectively, we can check for the absence of correlations between $V_{6L}$ and $V_3^2$, $V_{6L}$ and $V_2^3$, and $V_{7L}$ and $V_2^2V_3$ by checking the following equalities: 
\begin{eqnarray}
\label{A1}
\frac{\left\langle V_6 V_3^{*2} v_3^2 \right\rangle}
       {\left\langle V_6 V_3^{*2} \right\rangle \left\langle v_3^2 \right\rangle} & = &
\frac{\left \langle v_3^6 \right\rangle}
       {\left\langle v_3^4 \right\rangle \left\langle v_3^2 \right\rangle},
\\
\label{A2}
\frac{\left\langle V_6 V_2^{*3} v_2^2 \right\rangle}
       {\left\langle V_6 V_2^{*3} \right\rangle \left\langle v_2^2 \right\rangle} & = &
\frac{\left \langle v_2^8 \right\rangle}
       {\left\langle v_2^6 \right\rangle \left\langle v_2^2 \right\rangle},
\\
\label{A3}
\frac{\langle V_7 V_2^{*2} V_3^* v_2^2 \rangle}{\langle V_7 V_2^{*2} V_3^* \rangle\, \langle v_2^2 \rangle} 
      & = & \frac{\langle v_2^6 v_3^2 \rangle}{\langle v_2^4 v_3^2 \rangle\langle v_2^2 \rangle } =
\frac{\left \langle v_2^6 \right\rangle}
       {\left\langle v_2^4 \right\rangle \left\langle v_2^2 \right\rangle}.
\end{eqnarray}
The second equality in Eq.~(\ref{A3}) tests the statistical independence of fluctuations in the magnitudes $v_2$ and $v_3$ of the elliptic and triangular flow. Figs.~\ref{F9}a-c support the validity of all of these relations. The validity of these relations allows us to obtain the mode coupling coefficients $\chi_{633}$, $\chi_{6222}$ and $\chi_{7223}$ from Eqs.~(\ref{eq14}), (\ref{eq15}) and (\ref{eq18}). 

Using the decompositions (\ref{VlVnl_old}) to eliminate $V_{4L}$ and $V_{5L}$ and assuming the absence 
of correlations between the leading and mode coupling terms in the decompositions (\ref{chi_n}) for $V_6$ and $V_7$, we can similarly derive the following relations:
\begin{eqnarray}
\label{A4}
\frac{\left\langle (V_6 V_2^* V_4^* - \chi_{6222} \chi_{422} v_2^6) v_2^2 \right\rangle}
       {\left\langle V_6 V_2^* V_4^* - \chi_{6222} \chi_{422} v_2^6 \right\rangle \left\langle v_2^2 \right\rangle} & = & 
\frac{\left \langle v_2^4 \right\rangle}
       {\left\langle v_2^2 \right\rangle ^2 },
\\
\label{A5}
\frac{\left\langle (V_7 V_2^* V_5^* - \chi_{7223} \chi_{523} v_2^4 v_3^2) v_2^2 \right\rangle}
       {\left\langle V_7 V_2^* V_5^* - \chi_{7223} \chi_{523} v_2^4 v_3^2 \right\rangle \left\langle v_2^2 \right\rangle} & = &
\frac{\left \langle v_2^4 \right\rangle}
       {\left\langle v_2^2 \right\rangle ^2 },
\\
\label{A6}
\frac{\left\langle (V_7 V_3^* V_4^* - \chi_{7223} \chi_{422} v_2^4 v_3^2) v_3^2 \right\rangle}
       {\left\langle V_7 V_3^* V_4^* - \chi_{7223} \chi_{422} v_2^4 v_3^2 \right\rangle \left\langle v_3^2 \right\rangle} & = &
\frac{\left \langle v_3^4 \right\rangle}
       {\left\langle v_3^2 \right\rangle ^2 }.
\end{eqnarray}
Substituting the already validated relations (\ref{chi_old}) allows to rewrite these as
\begin{widetext}
\begin{eqnarray}
\label{A7}
&&\frac{\langle V_6 V_2^* V_4^* v_2^2 \rangle \langle v_2^4 \rangle \langle v_2^6 \rangle 
             - \langle V_6 V_2^{*3} \rangle \langle V_4 V_2^{*2} \rangle \langle v_2^8 \rangle}
           {\langle V_6 V_2^* V_4^* \rangle \langle v_2^2 \rangle \langle v_2^4 \rangle \langle v_2^6 \rangle 
             - \langle V_6 V_2^{*3} \rangle \langle V_4 V_2^{*2} \rangle \langle v_2^2 \rangle 
                \langle v_2^6\rangle} 
    = \frac{\left \langle v_2^4 \right\rangle}{\left\langle v_2^2 \right\rangle^2},
\\    
\label{A8}
&&\frac{\langle V_7 V_2^* V_5^* v_2^2 \rangle \langle v_2^2 v_3^2 \rangle \langle v_2^4 v_3^2 \rangle 
             - \langle V_7 V_2^{*2} V_3^* \rangle \langle V_5 V_2^* V_3^* \rangle \langle v_2^6 v_3^2 \rangle}
            {\langle V_7 V_2^* V_5^* \rangle \langle v_2^2 \rangle \langle v_2^2 v_3^2 \rangle 
             \langle v_2^4 v_3^2 \rangle 
              - \langle V_7 V_2^{*2} V_3^* \rangle \langle V_5 V_2^* V_3^* \rangle \langle v_2^2 \rangle 
              \langle v_2^4 v_3^2 \rangle} 
   = \frac{\left \langle v_2^4 \right\rangle}{\left\langle v_2^2 \right\rangle ^2},
\\
\label{A9}
&&\frac{\langle V_7 V_3^* V_4^* v_3^2 \rangle \langle v_2^4 \rangle \langle v_2^4 v_3^2 \rangle
             - \langle V_7 V_2^{*2} V_3^* \rangle \langle V_4 V_2^{*2} \rangle \langle v_2^4 v_3^4 \rangle}
            {\langle V_7 V_3^* V_4^* \rangle \langle v_3^2 \rangle \langle v_2^4 \rangle 
             \langle v_2^4 v_3^2 \rangle
             - \langle V_7 V_2^{*2} V_3^* \rangle \langle V_4 V_2^{*2} \rangle \langle v_3^2 \rangle 
             \langle v_2^4 v_3^2 \rangle} 
    = \frac{\left \langle v_3^4 \right\rangle}{\left\langle v_3^2 \right\rangle^2}.
\end{eqnarray}
\end{widetext}
Figures~\ref{F9}d,e show these relations (and thus the underlying assumptions) to be valid. 

%
%

\section{Resonance decay effects}
\label{appb}
\vspace*{-5mm}

All results shown in the main body of this paper were computed from the final charged hadron spectra, including all resonance decay contributions. Since this is numerically costly, we explore in this Appendix to what extent a simpler calculation that takes directly emitted positively charged pions as a proxy for all charged hadrons would have distorted the results. Fig.~\ref{F9} shows that the differences are generally small: only for the trilinear (and, to a lesser extent, for the bilinear) coupling contributions between elliptic and triangular flow to $V_7$ (panels f,h,F,H) do we observe significant corrections from kaons, protons, and resonance decay pions, mostly in non-central collisions. Generically, the inclusion of resonance decay pions and heavier stable hadrons tend to slightly increase the nonlinear flow coupling coefficients. Their importance is of the same order of magnitude for both of the initial condition models studied here. Due to the smallness of their effects, a calculation based on directly emitted pions alone would still have led us to the same conclusions that we have drawn from the full calculations presented in this paper.


\end{document}